\documentclass[%
 reprint,
superscriptaddress,
 amsmath,amssymb,
 aps,
]{revtex4-2}

\usepackage{natbib}
\usepackage[normalem]{ulem}
\usepackage{graphicx}
\usepackage{dcolumn}
\usepackage{bm}
\usepackage{siunitx}
\usepackage{hyperref}
\usepackage{orcidlink}
\usepackage{booktabs}
\usepackage[normalem]{ulem}

\hypersetup{
    colorlinks = true,
    linkcolor = blue,
    filecolor = magenta,      
    urlcolor = cyan,
    citecolor = blue
    }

\begin{document}

\preprint{APS/123-QED}

\title{Vortices in a two-dimensional fermionic superfluid}

\author{Artak Mkrtchyan \orcidlink{0009-0003-6873-481X}}
\email{artak.mkrtchyan@uni-hamburg.de}
\affiliation{Institute for Quantum Physics, Universität Hamburg, Luruper Chaussee 149, 22761 Hamburg, Germany}

\author{Hans Leonard Michel \orcidlink{0009-0009-6014-2469}}
\affiliation{Institute for Quantum Physics, Universität Hamburg, Luruper Chaussee 149, 22761 Hamburg, Germany}
\affiliation{The Hamburg Center for Ultrafast Imaging, Universität Hamburg,  Luruper Chaussee 149, 22761 Hamburg, Germany}

\author{Moritz von Usslar \orcidlink{0009-0001-7064-101X}}
\affiliation{Institute for Quantum Physics, Universität Hamburg, Luruper Chaussee 149, 22761 Hamburg, Germany}
\affiliation{The Hamburg Center for Ultrafast Imaging, Universität Hamburg,  Luruper Chaussee 149, 22761 Hamburg, Germany}

\author{René Henke \orcidlink{0000-0002-7075-3812}}
\affiliation{Institute for Quantum Physics, Universität Hamburg, Luruper Chaussee 149, 22761 Hamburg, Germany}
\affiliation{The Hamburg Center for Ultrafast Imaging, Universität Hamburg,  Luruper Chaussee 149, 22761 Hamburg, Germany}

\author{Cesar R. Cabrera \orcidlink{0000-0003-3867-9074}}
\affiliation{Institute for Quantum Physics, Universität Hamburg, Luruper Chaussee 149, 22761 Hamburg, Germany}
\affiliation{The Hamburg Center for Ultrafast Imaging, Universität Hamburg,  Luruper Chaussee 149, 22761 Hamburg, Germany}
\affiliation{Instituto Galego de F{\'{i}}sica de Altas Enerx{\'{i}}as,  Universidade de Santiago de Compostela, 
\\ Santiago de Compostela 15782,  Spain}

\author{Henning Moritz \orcidlink{0000-0002-4854-5806}}
\email{henning.moritz@uni-hamburg.de}
\affiliation{Institute for Quantum Physics, Universität Hamburg, Luruper Chaussee 149, 22761 Hamburg, Germany}
\affiliation{The Hamburg Center for Ultrafast Imaging, Universität Hamburg,  Luruper Chaussee 149, 22761 Hamburg, Germany}
\date{\today}

\begin{abstract}
    Quantized vortices are a hallmark of superfluidity. 
    They have proven to be an indispensable probe for systems ranging from superfluid helium to type-II superconductors.
    In ultracold two-dimensional (2D) Fermi gases, where phase fluctuations are enhanced and the Berezinskii-Kosterlitz-Thouless (BKT) mechanism governs the superfluid transition, their observation has remained elusive.
    Here, we report on the controlled creation of quantized vortices in a homogeneous 2D gas in the BEC--BCS crossover.
    By continuously tuning the velocity of two moving obstacles, we observe quantized steps in the absolute vortex charge.
    Matter-wave interferometry reveals the characteristic phase winding, quantized in units of $2\pi$.
    Measurements on 2D and 3D gases show a striking difference: only in 2D does the vortex number grow with temperature, reflecting thermal activation of BKT vortex pairs.
    Our approach provides evidence of fermionic superfluidity in 2D and offers a means to study BKT physics, vortex-controlled transport and resistance, as well as to detect unconventional superfluidity in spin-imbalanced 2D gases.
\end{abstract}

\maketitle

Superconductivity and fermionic superfluidity in two dimensions stand at the frontier of quantum many-body physics. 
Unlike in three-dimensional counterparts, true long-range order is suppressed by phase fluctuations \cite{Wagner_1966, Hohenberg_1967, Mermin_1968}.
Nevertheless, frictionless flow can still arise below the Berezinskii-Kosterlitz-Thouless (BKT) transition \cite{Berezinskii_1970, Kosterlitz_1973}, making 2D  the marginal dimension for superfluidity.
Remarkably, it is precisely in this fluctuation-dominated regime
where quasi-2D layered materials exhibit the highest critical temperatures among ambient-pressure superconductors \cite{Keimer2015-dj}. 
Understanding superfluidity in this correlation- and fluctuation-dominated regime remains a central challenge for theory, making a compelling case for quantum simulation.

\begin{figure}[t!]
    \centering
    \includegraphics[width = \columnwidth]{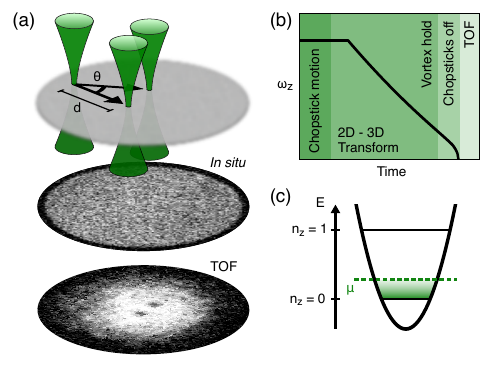}
    \caption{\textbf{Vortex creation and detection.} (a) An obstacle (green chopstick) is ramped up to a potential height of $V \approx \SI{0.6}{\mu}$, then moved $d \approx \SI{10}{\micro m}$ through the cloud while being split in two with an opening angle of $\theta = \SI{65}{\degree}$. Middle panel: an \textit{in situ} image of the cloud after motion and rampdown of chopsticks. The created vortices are so small that they cannot be seen \textit{in situ}. The image of the vortex pair in the bottom panel is obtained with the schematic detection procedure depicted in (b): 
    reduction of vertical trapping to transform to 3D within \SI{100}{ms} while ramping to the BEC regime. The chopsticks are then switched off, and the cloud is imaged after a \SI{3}{ms} TOF. Vortices then appear as clear density depletions in absorption. (c) The initial cloud is in the 2D regime as the chemical potential $\mu$ lies below the first excited state $\hbar \omega_{z}$ of the vertical confinement.}
    \label{figure: 1}
\end{figure}

\begin{figure*}[t!]
    \centering
    \includegraphics[width = 2\columnwidth]{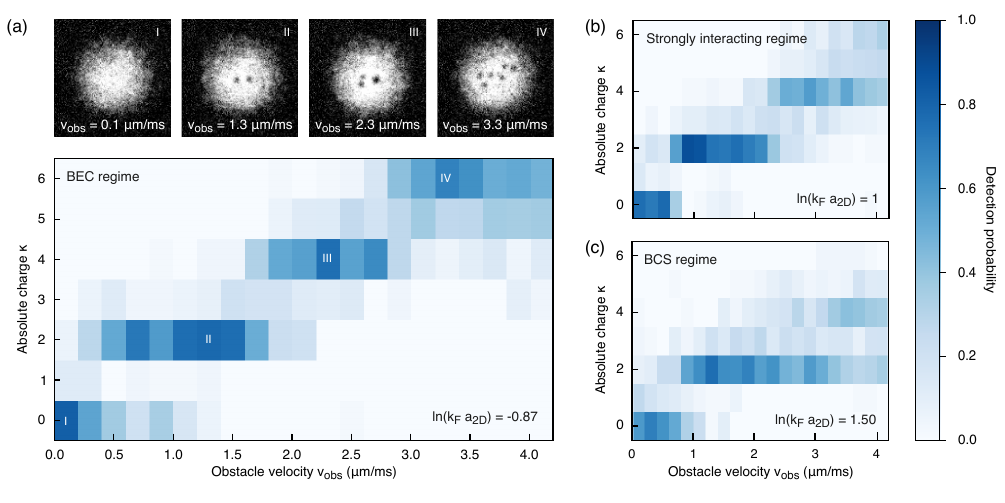}
    \caption{\textbf{Stepwise vortex creation.} (a) When continuously increasing the obstacle velocity, the absolute vortex charge $\kappa$ increases in steps of two at odd multiples of the onset velocity. Three steps can be accessed within our velocity range in the BEC regime at $\ln(k_{\mathrm{F}} a_{\mathrm{2D}}) = -0.87$. For each step, example single-shot absorption images are shown (see Roman numerals I-IV). At high velocities, doubly charged vortices with bigger cores can be created, as seen in III. (b), (c) Equivalent measurements in the strongly interacting and the BCS regime at $\ln(k_{\mathrm{F}} a_{\mathrm{2D}}) = 1$ and $\ln(k_{\mathrm{F}} a_{\mathrm{2D}}) = 1.50$, respectively. 50 shots were taken per obstacle velocity.}
    \label{figure: 2}
\end{figure*}

Ultracold Fermi gases have become a powerful platform for the quantum simulation of strongly correlated 2D systems.
Their exceptional tunability and local control have enabled direct probes of superfluidity through transport and critical-velocity measurements \cite{Esslinger_2012, Sobirey_2021}.
Yet quantum vortices - a hallmark signature historically used to identify superfluids \cite{Vinen_1961, Sanders_1969, Henini_1996, Cornell_1999, Dalibard_2000, Ketterle_2001, Ketterle_2005, Deveaud-Pledran_2008} - remain unobserved in 2D Fermi gases.
From superfluid helium to type-II superconductors and Bose-Einstein condensates, vortices are sensitive probes of superfluid properties: their core structure and their interactions are set by the coherence length and phase rigidity \cite{Donnelly_1991, Tinkham_1996, Fetter_2009, Ho_2006}, while their dynamics govern transport and dissipation \cite{Anderson_1966, Stephen_1965, Vinokur_1994, Siggia_1980, Campbell_2014, Roati_2018, Roati_2025}, with free vortices above the BKT transition leading to resistive transport \cite{Reppy_1978, Nelson_1979, Siggia_1980, Minnhagen_1987}.
Finally, they play a central role in nonequilibrium phenomena such as Kibble--Zurek dynamics \cite{Kibble_1976, Zurek_1985, Anderson_2008} and turbulent flow \cite{Vinen_1957, Neely_2019, Helmerson_2019, Shin_2019}.

Vortices have been studied extensively in 3D Bose and Fermi gases \cite{Cornell_1999, Dalibard_2000, Ketterle_2001, Ketterle_2005, Ketterle_2006, Fetter_2009}.
Planar vortex dynamics have also been probed in oblate gases \cite{Anderson_2010, Shin_2018, Gauthier_2019, Johnstone_2019, Pan_2021, Roati_2021, Pan_2022, Roati_2025}, which, however, remained kinematically 3D \footnote{That is, the chemical potential or the temperature exceeds the level spacing along the tightly confined direction.}.
In kinematically 2D gases, vortices have so far only been studied in the weakly interacting Bose regime \cite{Dalibard_2005, Shin_2013, Foot_2022, Foot_2023, Hung_2025}.
This gap reflects a detection challenge.
In 3D, vortex cores are magnified by hydrodynamic time-of-flight (TOF) expansion in the BEC regime until their density depletion becomes visible in absorption imaging \cite{Dalibard_2000, Ketterle_2001, Ketterle_2005}.
In strongly interacting 2D gases, however, this approach fails: TOF expansion along the previously tightly confined direction is too rapid for the core size to adjust \cite{Smith_1998, Modugno_2000, Dalibard_2005}, which, \textit{in situ}, is on the order of the interparticle spacing \cite{Ho_2006}. 
The phase fluctuations, native to 2D, are furthermore converted to density modulations in TOF \cite{Shin_2012, Mathey_2014, Foot_2025}, obscuring the small cores.

In this letter, we report the successful controlled creation and observation of vortices in a homogeneous 2D Fermi superfluid, where the detection challenges are addressed by converting the gas from 2D to 3D prior to TOF \cite{Shin_2013}.
By measuring quantized steps in the absolute vortex charge and quantized phase winding in multiples of $2 \pi$, we observe defining characteristics of quantum vortices.
We additionally observe that the number of detected vortices grows with increasing temperature, typical for thermal vortex activation in BKT systems.

The experiments are conducted in a gas of $^{6}\text{Li}$ atoms prepared in the lowest two hyperfine states \cite{Hueck_2018}.
The radial confinement is provided by a hard-wall ring potential with a radius of  $R \approx \SI{60}{\micro m}$, leading to a homogeneous density distribution as illustrated in Fig.~\ref{figure: 1}(a). 
Strong confinement along the vertical direction is achieved by loading the atoms into a single layer of a blue-detuned optical lattice, resulting in trapping frequencies of up to $\omega_{\mathrm{z}} \approx 2\pi \cdot \SI{20}{kHz}$.
This confinement freezes out the transverse degree of freedom when the chemical potential $\mu$ lies below the first excited state of the trapping potential (see Fig.~\ref{figure: 1}(c)).
In our experiments, this condition is met with $\mu / \hbar \omega_{\mathrm{z}} \lesssim 0.6$. 
The interaction strength between fermions is tuned via a Feshbach resonance, allowing us to probe the BEC–-BCS crossover, which is parametrized by the dimensionless interaction parameter $\ln(k_{\mathrm{F}} a_{\mathrm{2D}})$.
Here, $k_\mathrm{F} = \sqrt{4\pi n_{\mathrm{2D}}}$ is the Fermi wavevector, $n_{\mathrm{2D}}$ is the area density per spin state and $a_{\mathrm{2D}}$ is the two-dimensional scattering length \cite{supmat}.
For all experiments, the typical temperatures measured after an adiabatic ramp to the BEC regime ($\ln(k_{\mathrm{F}} a_{\mathrm{2D}}) \approx -0.9$) are $T^* / T_\mathrm{F} \lesssim 0.01$, where $T_\mathrm{F}$ is the Fermi temperature.

We produce vortices in a controlled manner across a wide range of interactions employing the so-called chopstick method \cite{Anderson_2016, Roati_2021, Hung_2025}.
Here, two repulsive initially overlapping obstacles are moved apart with a velocity $v$ and an opening angle $\theta$, as illustrated in Fig.~\ref{figure: 1}(a). 
The forward and opening motion of the obstacles induces flow, which leads to the nucleation of a vortex-antivortex pair at the positions of the obstacles when a onset velocity $v_{\mathrm{vc}}$ is exceeded \cite{Anderson_2016}. 
To address the aforementioned detection issues in 2D gases, we adapt a protocol  in which the system is adiabatically transformed to 3D by reducing the vertical confinement before being released from the trap \cite{Shin_2013, supmat}, shown in Fig.~\ref{figure: 1}(b). 
This mitigates detrimental effects from phase fluctuations and rapid vertical expansion beyond the depth of field of the imaging apparatus.

A typical image of the cloud above the vortex creation onset velocity is shown in Fig.~\ref{figure: 1}(a), where a vortex-antivortex pair can be seen at the final chopstick positions.
When further increasing the chopstick velocity, we observe an increase in vortex numbers.
We also find that some vortices feature substantially larger cores; see, e.g., Fig.~\ref{figure: 2}(a)-III, indicating that they carry two (or more) quanta of circulation \cite{Ketterle_2002, Ketterle_2004, Takahashi_2007, Hirano_2010, Roati_2021,Bagnato_2022}. 
When counting these large-core vortices as doubly charged \cite{supmat} and plotting the sum of the  absolute vortex charge $\kappa$ as a function of obstacle velocity in Fig.~\ref{figure: 2}(a-c), we observe two features that demonstrate a quantized response to a continuous tuning parameter. 
First, the absolute charge increases by two, as expected for the creation of vortex-antivortex pairs.
Second, the increase is stepwise, with each subsequent step located at $(2N + 1) \cdot v_{\mathrm{vc}}$, where $N = 0, 1, 2, ...$, reflecting a washboard excitation behavior \cite{Roati_2022}.
We observe this quantized vortex creation response across a wide range of interactions. 
As expected, a significant increase in the onset velocity from the BEC to the BCS regime is observed, in line with the larger speed of sound $c_s$ along the crossover \cite{Bohlen_2020}. 
The onset is always found at a similar fraction of $c_{\mathrm{s}}$ \cite{Anderson_2010, Shin_2015, Anderson_2016, Bohlen_2020, Roati_2021, Shin_2018} and hence doubles between $\ln(k_{\mathrm{F}} a_{\mathrm{2D}}) = -0.87$ and $\ln(k_{\mathrm{F}} a_{\mathrm{2D}}) = 1.50$ from $\approx \SI{0.6}{\micro m/ms}$ to $\approx \SI{1.2}{\micro m/ms}$.
As a consequence, an absolute charge of up to six can be created in the BEC regime, and only four in the BCS regime within our velocity range.

Only being able to detect vortices after the transformation to 3D raises the question of whether the induced flow in 2D is normal and merely projected onto quantized vortices during the transformation.
To rule out this scenario, we introduce a hold time of up to \SI{750}{ms} in 2D after vortex creation.
For a normal fluid this would cause an upward shift of the onset velocity as a result of viscous damping until the moment of projection.
This hold time far exceeds the damping timescale of $\tau \approx \SI{37}{ms}$, estimated for a normal fluid with minimal shear viscosity of order $\hbar n$ and our experimental system size and density \cite{Thomas_2011, Bohlen_2020, Zwierlein_2020, supmat}.
Since we measure no statistically significant increase in the onset velocity required for vortex creation for hold times up to \SI{750}{ms} (data shown in \cite{supmat}), we conclude that the 2D system is already superfluid and supports quantum vortices.

\begin{figure}[t!]
    \centering
    \includegraphics[width = \columnwidth]{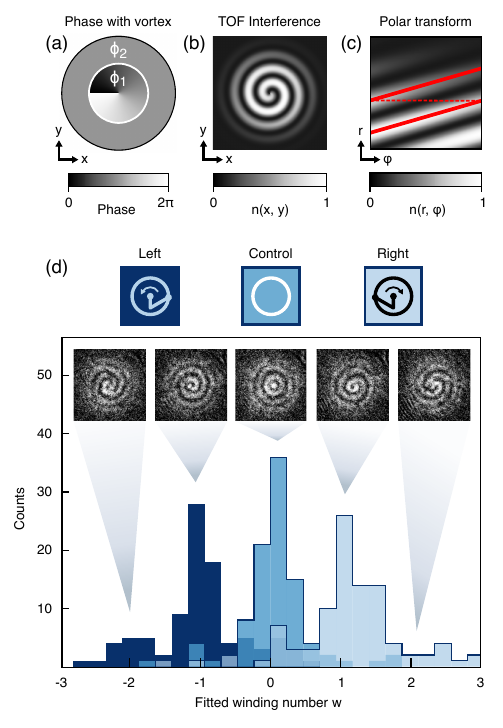}
    \caption{\textbf{Detection of phase winding.} (a) Schematic illustration of cloud preparation for interference. The cloud is separated into two concentric reservoirs: an inner disk with phase winding $\phi_{\mathrm{1}}$, provided by a charge-one vortex at the center, and an outer ring with a constant phase $\phi_{\mathrm{2}}$. (b) Interference of the reservoirs from (a) yields a spiral structure in TOF. (c) Transformation of the interference image from (b) into polar coordinates results in linear ridges from which the winding number $w$ can be extracted. (d) Histograms showing extracted winding numbers from interference patterns: the pictograms at the top illustrate the chopstick motion to generate a left- or right-rotating (dark or light-blue)  direction of circulation, and the corresponding histograms are peaked at winding numbers of $w = \pm 1$, alongside a control group with no chopstick motion (medium-blue) accumulating at $w = 0$. A small feature also appears at $w = -2$. The measurements depicted here are performed at $\ln(k_\mathrm{F} a_\mathrm{2D}) = 0.02$.} 
    \label{figure: 3}
\end{figure}

In addition to the quantized formation of vortices, we measure the phase and observe the characteristic $2\pi\cdot n$ phase winding of a superfluid containing a vortex.
While the phase of an atomic gas is not accessible in direct absorption imaging, relative phase differences can be observed via the interference with a reference cloud.
This is done by preparing two separated concentric reservoirs \cite{Roati_2022}, as depicted in Fig.~\ref{figure: 3}(a). The outer ring reservoir has a constant phase, whereas the inner disk contains a phase winding provided by a single vortex. 
The chopstick motion is modified to move one vortex with a fixed chopstick velocity of $\SI{0.8}{\micro m/ ms}$ into the center of the inner disk, while the other is moved into the separating barrier, disrupting its flow.
This is schematically illustrated in the pictograms of Fig.~\ref{figure: 3}(d).
Following a $\SI{10}{ms}$ hold time after vortex creation, we again use the detection sequence shown in Fig.~\ref {figure: 1}(b) and observe the interference pattern in TOF \cite{supmat}.
The expected interference pattern in the presence of a singly charged vortex is shown in Fig.~\ref{figure: 3}(b), while experimental images can be seen in Fig.~\ref{figure: 3}(d) for various chopstick motions. 

To quantitatively evaluate the phase winding from the interference patterns, the images are transformed to polar coordinates, where the patterns appear as a series of linear ridges, as illustrated in Fig.~\ref{figure: 3}(c).
The fitted slope of the ridges provides access to the winding number $w = 2\pi \cdot s/l$, where $s$ is the slope of the ridges and $l$ is the radial distance between the ridges \cite{supmat}.
Fig.~\ref{figure: 3}(d) shows the resulting histogram of the extracted winding numbers for an interaction strength of $\ln(k_{\mathrm{F}} a_{\mathrm{2D}}) = 0.02$.
The histograms for experimental sequences intended to induce left-handed, right-handed, or no circulation are shown in dark-, light-, and medium-blue, respectively.
The corresponding histograms show prominent peaks at windings $w = 1$ and $w=-1$, as expected for right- and left-rotating singly charged vortices with a phase winding of $2\pi$.
Naturally, where no circulation was induced, the histogram peaks at $w=0$.
A small feature also appears at $w = -2$, indicating the occasional formation of doubly charged vortices.
We recover similar histograms at interactions $\ln(k_{\mathrm{F}} a_{\mathrm{2D}}) = 1.08$ and $\ln(k_{\mathrm{F}} a_{\mathrm{2D}}) = 1.6$ (not shown), yet without significant $w = \pm 2$ contributions, consistent with a higher speed of sound relative to the excitation velocity.

\begin{figure}[h!]
    \includegraphics[width = \columnwidth]{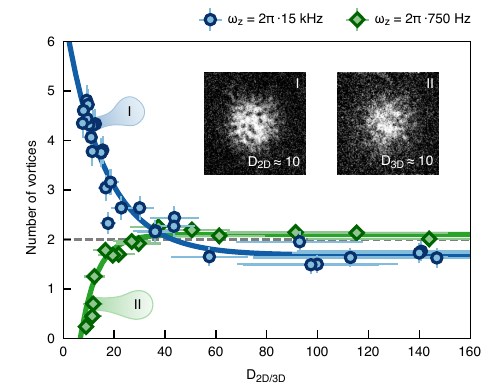}
    \caption{\textbf{Dimensionality-dependent thermal vortex activation.}
    As the gas is heated, additional vortices appear only in 2D (blue circles). 
    Their number grows exponentially with decreasing PSD (blue solid line), as expected for thermal activation of vortex-antivortex pairs below the BKT transition.
    In 3D (green diamonds, solid line as a guide to the eye), the vortex number decreases instead.
    In the coldest gases (high PSD), the chopstick motion at $\SI{0.8}{\micro m/ms}$ creates a single vortex--antivortex pair in both cases.
    The insets show single-shot images of the 2D and 3D cloud after the detection ramp.}
    \label{figure: 4}
\end{figure}

Finally, we examine how a 2D gas responds to heating and find that it behaves strikingly differently from a 3D gas probed in the same apparatus.
In 2D, the vertical trapping frequency is tuned to $\omega_{\mathrm{z}} = 2 \pi \cdot \SI{15}{kHz}$, whereas in 3D it is relaxed to $2 \pi \cdot \SI{750}{Hz}$, allowing the gas to occupy a few vertical oscillator levels.  
We parametrically heat the systems by modulating the power of the vertically confining lattice. 
After thermalization, the chopstick motion is performed in the strongly interacting regime at $\SI{832}{G}$ (corresponding to $\ln(k_{\mathrm{F}} a_{\mathrm{2D}}) \approx 1$ and $(k_{\mathrm{F}} a_{\mathrm{3D}})^{-1} \approx 0$)  at a fixed velocity of $\SI{0.8}{\micro m / ms}$.
We then count the number of vortices and plot them as a function of phase space density (PSD) with $D_{\textrm{2D}}=n_{\mathrm{2D}} \lambda^{2}$ and $D_{\mathrm{3D}}=n_{\mathrm{3D}} \lambda^{3}$, where $\lambda = h / \sqrt{2 \pi m_{\mathrm{d}} k_{\mathrm{B}} T^{*}}$ is the de Broglie wavelength and $m_{\mathrm{d}} = 2 \, m_{\mathrm{Li}}$ is the dimer mass \footnote{$n_{\mathrm{3D}}$ is estimated from the Thomas-Fermi profile at \SI{750}{G}.}\footnote{We note that $D_{\mathrm{2D}}$ and $D_{\mathrm{3D}}$ do not directly represent the PSDs at \SI{832}{G}, where the vortices are created, since the corresponding temperatures $T^*$ are measured after an adiabatic ramp to the BEC regime.
Thus, these quantities merely serve to indicate the degree of heating.}.

As seen in Fig. 4, in 2D the mean number of detected vortices grows rapidly with decreasing PSD, whereas it drops in 3D.
We find that this growth is well described by an exponential dependence on the phase space density, the behavior expected for thermally activated vortex--antivortex pairs below the BKT transition
\cite{Dalibard_2006, Shin_2013, Foot_2022, Timm_1996}.
The disappearance of thermal vortex activation when only a few vertical oscillator levels are occupied ($ \mu / \hbar \omega_{\mathrm{z}} \approx 4$) highlights the strong sensitivity of thermal behavior to dimensionality. 
This underscores the importance of reaching the kinematically 2D regime to observe the defining features of 2D superfluidity, which are central to systems such as layered high-temperature superconductors and thin helium films.

In summary, we established the controlled creation and observation of quantum vortices in a two-dimensional Fermi gas.
Quantized steps in the vortex charge and a $2\pi$ phase winding identify the excitations as quantum vortices.
The absence of measurable damping with hold time confirms that they are carried by a superfluid already present in 2D, and their exponential thermal activation, absent in 3D, reveals the BKT character of the transition.
We find that the extremely low temperatures achieved here are crucial for the deterministic generation and manipulation of vortices in 2D.
First measurements indicate that these methods can be extended to spin-imbalanced 2D gases, establishing vortices as a general probe of 2D fermionic superfluidity.
This is of particular interest in homogeneous 2D systems, where exotic phases like the FFLO phase are thought to be particularly stable \cite{Ferrell_1964, Ovchinnikov_1965}. 
Finally, the precise positioning of vortices enables the study of vortex-mediated resistance in a 2D channel \cite{Esslinger_2026}.

\textbf{Acknowledgments.}
We gratefully acknowledge fruitful discussions with H.\ Biss, G.\ Del Pace, T.\ Giamarchi, T.\ Enss, J.\ Levinsen, and M.\ Parish.
This work is supported by the Deutsche Forschungsgemeinschaft (DFG, German Research Foundation) in the framework of SFB-925—project 170620586—and the Cluster of Excellence “CUI: Advanced Imaging of Matter”—EXC 2056—project ID 390715994 and co-financed by ERDF of the European Union and by “Fonds of the Hamburg Ministry of Science, Research, Equalities and Districts (BWFGB)”.
H. L. M. acknowledges support as a recipient of a grant from Wübben Wissenschaftsstiftung gGmbH.
C.R.C. acknowledges support from the European Research Council under project DIPOLAR$\textunderscore$SF (101222586) and by the Maria de Maeztu excellence unit grant CEX2023-001318-M funded by MICIU/AEI/10.13039/501100011033.

\newpage

\bibliography{references}

\newpage

\begin{center}
\textbf{\large Supplementary Material: Vortices in a 2D Fermionic Superfluid}
\end{center}

\setcounter{figure}{0}
\renewcommand{\thefigure}{S\arabic{figure}}

\section{Experimental plattform}
\subsection{Preparation of a homogeneous 2D Fermi gas}
The experiments are performed in a balanced spin-mixture of $^6$Li, prepared in the two lowest hyperfine states $| F, m_{F}\rangle = | \frac{1}{2}, \frac{1}{2} \rangle$ and $| \frac{1}{2}, -\frac{1}{2} \rangle$. 
Pre-cooled atoms are first loaded into a hybrid trap consisting of a highly elliptical red-detuned beam and a magnetic trap that provides radial confinement, while also serving as an offset field to tune the interparticle interactions.
After forced evaporative cooling, the atoms are transferred into a circular box trap of radius $R = \SI{60}{\micro m}$ in which the density is homogeneous.
Radial confinement is achieved with a blue-detuned ring barrier, generated by a set of axicons and projected onto the atoms by a high-resolution (NA = 0.62) microscope objective.
The cloud is then loaded into a single layer of a highly tunable blue-detuned \SI{532}{nm} accordion lattice, providing a vertical trapping frequency of up to $\omega_{\mathrm{z}} \approx 2 \pi \cdot \SI{20}{kHz}$.
Final evaporation through the ring barrier allows us to control the density of the atoms.
This results in the trap level spacing $\hbar \omega_{\mathrm{z}}$ being comparable to the Fermi energy $E_{\mathrm{F}}$ of the gas, allowing us to parametrize the system in terms of pure 2D quantities; see below.
The Fermi energy of a 2D gas with density $n_{\mathrm{2D}}$ then takes the form $E_{\mathrm{F}} = \hbar^{2} k_{\mathrm{F}}^{2}/2m$, where $k_{\mathrm{F}} = \sqrt{4 \pi n_{\mathrm{2D}}}$ is the Fermi momentum and $m$ is the mass of the atomic species.
At this point, the temperature of the gas after an adiabatic ramp to the BEC regime lies in the single-digit nanokelvin range, such that $k_B T^{\ast}\ll\hbar \omega_{\mathrm{z}}$.
We determine the temperature $T^{\ast}$ by mapping the momentum distribution to density using a matter-wave focusing technique \cite{Sobirey_2021}.

\subsection{2D Condition and scattering}
Expressing the 2D scattering length $a_{\mathrm{2D}}$ in terms of the 3D scattering length $a_{\mathrm{3D}}$, the harmonic oscillator length of the confining potential $l_{\mathrm{z}} = \sqrt{\hbar / (m \omega_{\mathrm{z}})}$ and the chemical potential of the gas $\mu$ yields

\begin{equation}
    a_{\mathrm{2D}} = l_{\mathrm{z}} \sqrt{\frac{\pi}{A}} e^{- \sqrt{\frac{\pi}{2}}l_{\mathrm{z}}/a_{\mathrm{3D}}} e^{-\frac{1}{2} \Delta w (\frac{\mu}{\hbar \omega_{\mathrm{z}}})}
\end{equation}

\noindent where $A \approx 0.905$ is a coefficient obtained by comparison to the exact 2D scattering theory, and $\Delta w$ is a correction function accounting for the fermionic character of the particles \cite{Petrov2001-kn,Levinsen2015-du,Turlapov2017-nl}.
We use $a_{\mathrm{2D}}$ to define the dimensionless interaction strength $\ln(k_{\mathrm{F}} a_{\mathrm{2D}})$, with which we parametrize the BEC-BCS crossover in 2D.

We consider the system to be truly 2D when $\mu / \hbar \omega_{\mathrm{z}} \lesssim 1$.
To meet this condition, we adjust the vertical trapping frequency $\omega_{\mathrm{z}}$ as the chemical potential $\mu$ changes across the BEC-BCS crossover at constant density.
In the experimental range from $\ln(k_{\mathrm{F}} a_{\mathrm{2D}}) = -0.87$ (BEC) to $\ln(k_{\mathrm{F}} a_{\mathrm{2D}}) = 1.76$ (BCS), the trapping frequency lies between $\SI{10}{kHz} \lesssim \omega_{\mathrm{z}} / 2\pi \lesssim \SI{20}{kHz}$, ensuring $\mu / \hbar \omega_{\mathrm{z}} \lesssim 0.6$ for all our 2D experiments.
We refer to \cite{Shi2015} to calculate the chemical potential at any given interaction strength.

\subsection{Realization of a 3D system}
A 3D system, where the gas occupies a few vibrational levels of the vertical confinement, is realized by reducing the trapping frequency of our accordion lattice.
Operating the lattice at low power in its open configuration yields a trap frequency of $\omega_{\mathrm{z}} = 2 \pi \cdot \SI{750}{Hz}$, while maintaining a comparable atom number.
At this weak trapping, residual magnetic-field confinement, used to control the scattering length, increases the density at the cloud center.

In the strongly interacting regime at \SI{832}{G}, we estimate the mean chemical potential in the cloud.
For this, we consider a Thomas-Fermi profile of a unitary Fermi gas along the vertical direction

\begin{equation} \label{equation: UFG chemical potential 3D}
    n_{\mathrm{3D, 832}} (z) = \frac{1}{3 \pi^{2}} \left( \frac{2 m}{\xi \hbar^{2}} \right)^{\frac{3}{2}} \left(\mu_{\mathrm{3D, 832}} - \frac{1}{2} m \omega_{\mathrm{z}}^{2} z^{2} \right)^{\frac{3}{2}}
\end{equation}

\noindent where $\xi = 0.37$ is the Bertsch parameter, $\mu_{\mathrm{3D, 832}}$ is the 3D chemical potential, and $z$ is the vertical spatial coordinate. 
For our experimental parameters, $\mu_{\mathrm{3D, 832}} / \hbar \omega_{\mathrm{z}} \approx 2 - 4$ for different degrees of heating, where lower values come from the loss of atoms as a result of greater heating.
Corresponding densities equate to $n_{\mathrm{3D, 832}}(z = 0) \approx 0.2 - \SI{0.5}{\micro m^{-3}}$.

\subsection{Observation of Vortices: from 2D to 3D} 
Detecting vortices in a 2D fermionic superfluid is particularly challenging compared to 3D systems. 
In both geometries, the in situ vortex core size is on the order of the healing length, which is typically comparable to the resolution of the imaging system.
Therefore, hydrodynamic time-of-flight (TOF) expansion is required to magnify the core. 
In 3D, the expansion is typically hydrodynamic for a substantial amount of time and preserves the density signatures of vortices. 
In 2D, however, enhanced phase fluctuations inherent to the BKT regime cause interference in TOF and give rise to density speckles that hide the vortex cores.
Moreover, removing the strong vertical confinement leads to rapid, non-hydrodynamic vertical expansion that washes out vortex contrast. 
Together, these effects make it challenging to reliably identify vortices in 2D fermionic systems. 
To overcome these limitations, we developed a careful detection protocol that preserves density and phase information associated with quantized circulation. 
Following vortex creation, the vertical confinement is adiabatically ramped down, transforming the system into a quasi-3D geometry.
This way, the dimensional crossover suppresses the BKT phase fluctuations, avoiding distortions during expansion, and rapid vertical expansion in TOF is mitigated, considerably increasing the contrast of the vortex cores.

Vortex detection further depends on the interaction regime. 
On the BEC side of the crossover, vortices produce clear density depletions and can be observed directly after a short ToF expansion. 
In the strongly interacting and BCS regimes, however, the reduced condensate fraction leads to weak density contrast. 
To enable reliable detection in these regimes, we adiabatically ramp the magnetic field into the BEC regime prior to imaging, thereby converting fermionic pairs into tightly bound molecules while preserving the phase structure of the superfluid. 
The cloud is then released and expands for 3 ms before absorption imaging, during which vortices appear as well-defined density depletions, free of speckles, allowing their unimpaired detection.
Residual radial magnetic field curvature decreases the radial cloud size during ToF and increases the signal-to-noise ratio.

\subsection{Chopstick generation and calibration}
The chopstick obstacles are projected into the cloud using the Texas Instruments Lightcrafter DLP 6500 digital micromirror device (DMD).
With a chopstick size of $D_{\mathrm{obs}} \approx \SI{6.5}{\micro m}$ and a step size of $0.06 \cdot D_{\mathrm{obs}}$ per frame, the frame rate range of $\approx \SI{150}{}$ - $\SI{6300}{Hz}$ provides access to chopstick velocities between $\SI{0.1}{}$ - $\SI{4.1}{\micro m /ms}$ that were used in this work.

The potential height of the chopsticks was calibrated by projecting a dimple into the cloud, at the same power as the chopsticks, and measuring the ratio of the density inside and outside the dimple.
This calibration was conducted in the BCS regime at $\ln(k_{\mathrm{F}} \, a_{\mathrm{2D}}) = 1.76$, where the chemical potential is comparable to the Fermi energy of a non-interacting ideal Fermi gas $E_{\mathrm{F}}$.
At usual operating densities of $n_{\mathrm{2D}} \approx \SI{1.8}{\micro m^{-2}}$, the potential height of the dimple is on the order of $V_{\mathrm{Dimple}} \approx 0.6 \, E_{\mathrm{F}}$.

\section{Measurement of onset velocity with hold time and damping estimation.}
\begin{figure}[h!]
    \centering
    \includegraphics[width = \columnwidth]{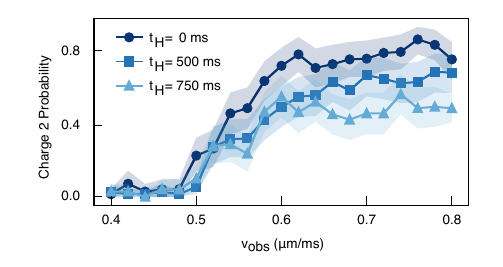}
    \caption{\textbf{Dissipationless flow in 2D} Probability to detect a total charge of two for hold times of \SI{0}{ms}, \SI{500}{ms} and \SI{750}{ms} after chopstick motion in 2D. While the probability decreases with increasing hold time, the onset velocity of \SI{0.5}{\micro m / ms} remains unchanged, suggesting that 2D flow is dissipationless.} 
    \label{figure: hold}
\end{figure}

Only being able to detect vortices in 3D raises the question of whether the induced flow in 2D is normal and is merely projected onto quantized vortices during the transformation to 3D.
To rule out this scenario, we introduce a hold time of up to \SI{750}{ms} in 2D after vortex creation.
Assuming a normal fluid, this would shift the onset velocity for vortex creation upward due to viscous damping.
Fig. \ref{figure: hold} shows the probability of observing an absolute vortex charge of two as a function of chopstick speed for different hold times at $\ln(k_{\mathrm{F}} a_{\mathrm{2D}}) = -0.03$.
The key observation is that no statistically significant increase in the onset velocity ($\approx \SI{0.5}{\micro m/ms})$ is registered for increasing hold times. 
This strongly indicates the absence of viscous damping in our measurements, in view of the longest damping time scale for a normal fluid of order \SI{37}{ms} estimated below.
We do see a reduction of charge-two probability at higher velocities, which, however, can be attributed to unpinning and annihilation of vortices the longer they are held. 

In the following, we set out to estimate the damping timescale of a vortex-antivortex pair in a normal fluid.
Our approach entails expanding the flow field into angular harmonics and deriving the decay time of the longest-lived mode, which serves as an upper limit.
This ansatz leans on three key assumptions:

\begin{enumerate}
    \item The fluid is incompressible, as the flow induced by a vortex-antivortex pair is slow with respect to the speed of sound.
    \item The minimum shear viscosity of a normal fluid is on the order of $\eta \sim \hbar n$, the quantum limit derived from kinetic theory \cite{Schaefer_2009, Thomas_2011}.
    \item The contact between the fluid and the trap walls is frictionless.
\end{enumerate}

The starting point for the timescale estimation is the scalar vorticity equation for an incompressible two-dimensional flow

\begin{equation} \label{equation: vorticity with omega}
    \partial_{t} \omega = \nu \nabla^{2} \omega \, ,
\end{equation}

\noindent where $\vec{\omega} = \nabla \times \vec{u} = (0, 0, \omega)$ is the vorticity of a velocity field $\vec{u}$ \cite{Batchelor_2000}.
Dissipation enters via the kinematic viscosity $\nu = \eta / \rho$, with the minimum shear viscosity $\eta \approx \hbar n$ \cite{Thomas_2011, Bohlen_2020} and the mass density $\rho = mn$.
The vorticity can be expressed in terms of a scalar stream function $\psi(r, \varphi, t)$ as $\omega = - \nabla^{2} \psi$, where streamlines of the fluid follow contours of constant $\psi$ \cite{Batchelor_2000}.
Using the stream function, Eq. \ref{equation: vorticity with omega} can be written as

\begin{equation} \label{equation: vorticity with psi}
    \partial_{t} \nabla^{2} \psi = \nu \nabla^{4} \psi \, .
\end{equation}

When the boundary conditions of an impermeable wall $\psi(R, \varphi) = 0$ and free slip $\frac{\partial^{2}}{\partial r^{2}}\psi - \frac{1}{r} \frac{\partial}{\partial r} \psi = 0$ at $r = R$ are imposed \cite{Batchelor_2000, Landau_1987}, a solution can be found via expansion into angular harmonics \cite{NIST:DLMF}

\begin{equation}
    \psi_{lj}(r, \varphi, t) = (A \, J_{l}(k_{lj} \, r) + B \, r^{l}) \, cos(l\varphi) \, e^{-\nu\frac{  \lambda_{lj}^{2}}{R} t} \, .
\end{equation}

\noindent Here, $J_{l}$ are Bessel functions of the first kind, $l = 0, 1, 2, ...$ and $j = 1, 2, ...$ are angular and radial modes respectively and $A$ and $B$ are constants.
Each mode decays on a timescale $\tau_{lj} = R^{2} / \nu \lambda_{lj}^{2}$, where $\lambda_{lj}$ is a dimensionless number for $lj$-mode arising from the solution with the specific boundary conditions.
In view of the experimentally realized vortex-antivortex pairs, where atoms flow in one direction between the pairs and back along the walls, the slowest decaying mode is the dipole mode with $l = 1, j=1$, yielding $\lambda_{11} = 3.054$.
Using $R \approx \SI{60}{\micro m}$ and $\nu = \hbar / m_{\mathrm{Li}} \approx \SI{1e-8}{m^{2}s^{-1}}$ yields a decay time of $\tau_{11} \approx \SI{37}{ms}$.
In the real experiment, higher modes contribute, making damping even faster.

Even when assuming the universal lower bound for the shear viscosity conjectured by Kovtun, Son and Starinets $\eta \geq \hbar n \frac{s}{4 \pi k_B}$ \cite{Starinets_2005} and using the entropy per particle measured in 3D at the critical point of $s \approx 0.73\cdot k_{\mathrm{B}}$ \cite{Zwierlein_2012}, the decay time $\tau_{11, \, \mathrm{KSS}} \approx \SI{640}{ms}$ is still shorter than the maximum hold time used in the experiment.

\section{Data analysis methods}
\subsection{Vortex counting}

\subsubsection{Image preparation and blob detection}
\begin{figure}[h]
    \centering
    \includegraphics[width = \columnwidth]{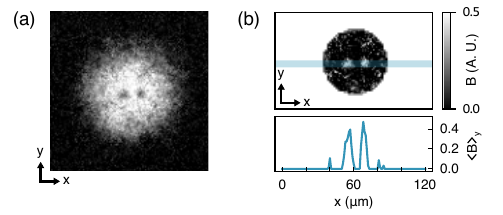}
    \caption{\textbf{Image processing for vortex detection.} (a) Original vortex-antivortex image. (b) Same image after clipping, background removal, and masking. The lower panel shows the brightness in the blue-shaded region, averaged along the y-direction.}
    \label{figure: masking}
\end{figure}

In order to analyze the positions and sizes of the vortices, they are isolated from the image by removing the density envelope of the cloud.
We do this by subtracting the original image from a Gaussian-filtered version, so the vortices appear as bright features on a dark background.
Clipping the brightest and darkest values of the original image beforehand reduces fine-grained noise, improving vortex visibility after background removal.
A circular mask excludes the low-density edge of the cloud from the region of interest, as vortex detection is expected to be unreliable and increasingly faulty there.
Fig. \ref{figure: masking} shows a vortex image before and after clipping, background removal, and masking.
To identify vortex position and size, we resort to a blob-detection function based on calculating the Laplacian of Gaussians and identifying the strongest responses. 

\subsubsection{Size thresholding}
\begin{figure}[h]
    \centering
    \includegraphics[width = \columnwidth]{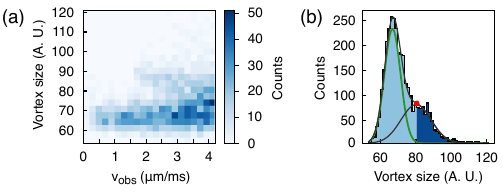}
    \caption{\textbf{Vortex core size thresholding.} (a) Vortex core size as a function of chopstick velocity at $\ln(k_{\mathrm{F}} a_{\mathrm{2D}}) = -0.87$. Above $v_{\mathrm{obs}} \sim \SI{1.5}{\mu m/ms}$ vortices with substantially larger cores are created. (b) Histogram of vortex core sizes integrated over chopstick velocity. Large-core vortices form a shoulder to the main size peak. The histogram is fitted with two Gaussians. Any vortex with core larger than $3 \sigma$ of the main peak (green Gaussian) is counted as charge two (dark region).}
    \label{figure: thresholding}
\end{figure}

An analysis of the vortex size distribution reveals a bimodal distribution.
It features a prominent peak for charge-one vortices and a shoulder at greater sizes for charge-two vortices.
To evaluate quantized charge creation correctly, the different vortex core sizes must be accounted for.
We fit the bimodal size distribution with two Gaussians and count all vortices with core sizes larger than the $3 \sigma$ range of the charge-one peak as doubly charged.
Fig. \ref{figure: thresholding} shows the velocity-resolved and velocity-integrated vortex core size distribution.

\subsection{Winding number extraction - spiral analysis}
To extract winding numbers from the imaged spirals, we developed an automated workflow that analyzes large datasets containing thousands of images.

The method relies on unwrapping the annular interference patterns in polar coordinates and analyzing the slopes of the resulting ridges.
A series of concentric rings transforms into parallel, flat ridges, whereas a spiral transforms into ridges with a finite slope, which is directly related to the winding number.
Because the interference patterns arise from well-defined, quantized phase fields, extracting the winding number from the ridge slope in polar coordinates constitutes a robust method of analysis.

\subsubsection{Finding an initial centre}
\begin{figure}[h]
    \centering
    \includegraphics[width = \columnwidth]{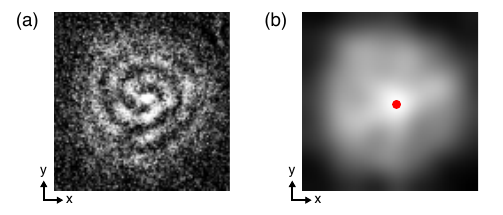}
    \caption{\textbf{Initial guess for the spiral center.} (a) Original spiral image. (b) Same image after smoothing with a Gaussian blur. The initial center guess is taken as the position of highest density in (b), marked in red.}
    \label{figure: supmat_initial_guess}
\end{figure}

The straightness of the ridges is highly sensitive to the center position from which the structure is unwrapped: if the proper center is not found, the ridges distort, making slope extraction unreliable.
Identifying the center that yields the least-distorted ridges therefore requires an iterative approach.
To start the iterative process, an initial guess for the center is made.
We found that smoothing the image with a Gaussian blur and using the center of the broad density envelope serves well for this initial guess.
Fig. \ref{figure: supmat_initial_guess} shows an original spiral image and its smoothed counterpart, with the position of highest density marked in red.

\begin{figure}[h]
    \centering
    \includegraphics[width = \columnwidth]{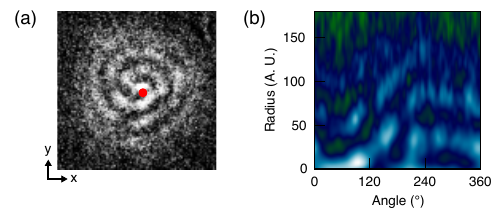}
    \caption{\textbf{Polar transform.} (a) Original spiral image with initial guess center marked in red. (b) Polar transform of the spiral image from the initial guess center.}
    \label{figure: supmat_initial_guess_polar}
\end{figure}

Following the initial guess, the image is transformed into polar coordinates. This is shown in Fig. \ref{figure: supmat_initial_guess_polar}.

\subsubsection{Ridge detection and tracking} 
\begin{figure}[h]
    \centering
    \includegraphics[width = \columnwidth]{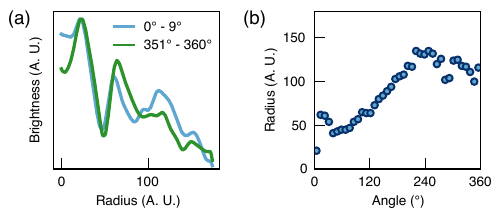}
    \caption{\textbf{Ridge tracking.} (a) The polar transform is subdivided into many angular bins, each yielding a radial density trace. The blue (green) trace is for the first (last) bin. (b) The ridge radius is found by tracking the maxima in (a).}
    \label{figure: supmat_initial_guess_ridge}
\end{figure}

To track the ridges, the polar image of a slightly blurred image is subdivided into narrow angular bins, and for each angular bin the density as a function of radius is calculated; see traces in Fig. \ref{figure: supmat_initial_guess_ridge} (a).
The peaks correspond to the interference maxima and are located using a peak-finding function.
The ridge is tracked across the full angular range by linking, bin by bin, the peaks closest in radius; the result is shown in Fig. \ref{figure: supmat_initial_guess_ridge} (b).

\subsubsection{Slope fitting and center position optimization}
\begin{figure}[h!]
    \centering
    \includegraphics[width = \columnwidth]{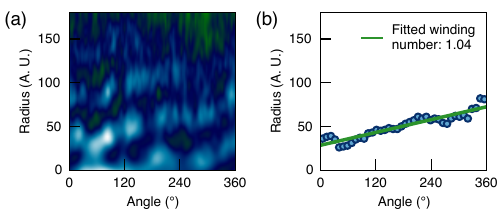}
    \caption{\textbf{Polar transform and slope extraction with optimized center.} (a) same as \ref{figure: supmat_initial_guess_polar} (b) but with optimized center for polar transform. Straight brightness ridges now allow for accurate slope extraction. (b) Brightness ridge fitted with a linear function to extract the slope from which the winding number can be determined.}
    \label{figure: supmat_optimal_polar}
\end{figure}

The slope of the ridge is determined using linear regression.
To reduce distortion of the ridges, the center position for the polar transform is optimized by minimizing the fit error.
Fig. \ref{figure: supmat_optimal_polar} shows the polar transform and corresponding ridge evaluated for the optimal center position.

The winding number $w$ is determined from the fitted slope $s$  via

\begin{equation} \label{equation: slope}
    s = \frac{\Delta r}{\Delta \phi} = \frac{wd}{2\pi}
\end{equation}

\noindent where $\Delta r$ is the radial extent of the ridge across the whole azimuthal angle $\Delta \phi = 2 \pi$ and $d$ is the radial distance between individual ridges.

\subsection{2D and 3D PSDs}
To compare the thermal behavior of 2D and 3D gases, the number of vortices is plotted against PSD for both cases (see Fig. 4 in the main text). 
Defined as 

\begin{equation}
    D_{\mathrm{2D}} = n_{\mathrm{2D}} \, \lambda^{2}
\end{equation}
\begin{equation}
    D_{\mathrm{3D}} = n_{\mathrm{3D}} \, \lambda^{3}
\end{equation}

\noindent respectively, $n_{\mathrm{2D}}$ and $n_{\mathrm{3D}}$ are not the densities in the strongly interacting regimes ($ln(k_{\mathrm{F}} \, a_{\mathrm{2D}}) = 1$ and $1 / k_{\mathrm{F}} \, a_{\mathrm{3D}}$ at $B = \SI{832}{G}$), where the experiments have been performed, but are estimated for the BEC regime where the measurement of temperature is possible ($B = \SI{750}{G}$).
As such, $D_{\mathrm{2D}}$ and $D_{\mathrm{3D}}$ are not the true PSDs but merely serve to indicate the degree of heating.

\subsubsection{2D BEC density estimation}
Due to the strong compression of the cloud in 2D, the measured density does not change across the BEC-BCS crossover such that $n_{\mathrm{2D, 750}} \approx n_{\mathrm{2D, 832}}$.

\subsubsection{3D BEC density estimation}
We consider the Thomas-Fermi profile of a BEC along the vertical direction

\begin{equation} \label{equation: 3D BEC density}
    n_{\mathrm{3D, 750}}(z) = \frac{\mu_{\mathrm{3D, 750}}}{g} (1 - \frac{z^{2}}{R_{\mathrm{z}}^{2}}) \, ,
\end{equation}

\noindent where $g$ is the dimensionless interaction parameter and $R_{\mathrm{z}}$ is the Thomas-Fermi radius.
From Equation \ref{equation: 3D BEC density} we determine the the central density for $z = 0$ from the measured area density $n_{\mathrm{2D}}$

\begin{equation} \label{equation: 3D BEC central density from 2D}
    n_{\mathrm{3D,750}}(z = 0) = \frac{3}{4} n_{\mathrm{2D}}^{2/3} \left( \frac{2}{3} \frac{m_{\mathrm{D}} \omega_{\mathrm{z}}^{2}}{g} \right)^{1/3} \, .
\end{equation}

\noindent Evaluating Equation \ref{equation: 3D BEC central density from 2D} for our experimental parameters results in densities of $n_{\mathrm{3D, 750}} \approx 0.4 - \SI{1.0}{\micro m^{-3}}$, roughly double of the true densities.
This difference, however, has no effect on the qualitative interpretation of Fig. 4.

\setlength{\tabcolsep}{12 pt}
\renewcommand{\arraystretch}{1.5}

\begin{table*}[b]
    \centering
    \begin{tabular}{c|cccccc} \toprule
        Figure   & $n_{\mathrm{2D}}$ ($\SI{}{\micro m}^{-2}$) & $B$ (G) & $\omega_{\mathrm{z}}/2 \pi$ (kHz) & $\mu$ (kHz) & $\ln(k_{\mathrm{F}} \, a_{\mathrm{2D}})$ \\ \midrule
        2(a)                       & $1.6 \pm 0.1$                                                & 750 & $\approx 10$   & $2.2 \pm 0.2$                         & $-0.87 \mp 0.01$ \\ 
        2(b)                       & $1.8 \pm 0.1$                                                & 832 & $\approx 15$   & $8.0 \pm 0.5$                         & $1.00 \mp 0.03$  \\ 
        2(c)                       & $1.8 \pm 0.1$                                                & 870 & $\approx 20$   & $9.7 \pm 0.6$                         & $1.50 \mp 0.03$ \\ 
        3(d)                       & $1.3 \pm 0.1$                                                & 780 & $\approx 10$   & $3.3 \pm 0.2$                         & $0.02 \mp 0.01$ \\ 
        4(2D)                      & $\approx 1.2 - 1.8$                                          & 832 & $\approx 15$   & $\approx 5.6 - 8.1$                   & $\approx 1.12 - 0.99$ \\ \midrule
        $\text{4(3D)}^{\ast}$      & $n_{\mathrm{3D, 823}} \approx 0.2 - \SI{0.5}{\micro m^{-3}}$ & 832 & $\approx 0.75$ & $\mu_{\mathrm{3D}} \approx 1.6 - 3.0$ & $1/k_{\mathrm{F}} \, a_{\mathrm{3D}} \approx 0$  \\
        $\text{4(3D)}^{\ast \ast}$ & $n_{\mathrm{3D, 750}} \approx 0.4 - \SI{1.0}{\micro m^{-3}}$ & -   & -              & -                                     & -  \\
        \bottomrule
    \end{tabular}
    \caption{\textbf{Experimental parameters.} Errors on the listed quantities are dominated by the systematic uncertainty on the density of $\approx 7 \, \%$. Statistical uncertainties enter on the order of $\approx 2-3 \, \%$. $\ast$ Estimation of true 3D densities using Equation \ref{equation: UFG chemical potential 3D} for a unitary Fermi gas. $\ast \ast$ Estimation of 3D densities for the evaluation of 3D PSDs in the BEC regime using Equation \ref{equation: 3D BEC central density from 2D}.}
    \label{table:parameters}
\end{table*}

\end{document}